\documentclass[12pt]{article}
\usepackage{epsfig}
\usepackage{mathtools}
\usepackage{amsfonts}
\usepackage{amssymb}
\usepackage{amsmath}
\usepackage{shuffle} 
\usepackage{color}
\usepackage{slashed}
\usepackage{cite}
\usepackage{caption}
\usepackage{subcaption}
\usepackage[]{breqn}

\usepackage{geometry} 
\usepackage{graphicx}
\graphicspath{{pictures/}}

\usepackage{multirow}           
\usepackage{array}              

\def\ep  {\varepsilon}

\begin{document}

\begin{center}

\vspace{3cm}

{\bf \Large Integral representation for three-loop banana graph} \vspace{1cm}

{\large M.A. Bezuglov$^{1,2,3}$}\vspace{0.5cm}

{\it $^1$Bogoliubov Laboratory of Theoretical Physics, Joint
    Institute for Nuclear Research, Dubna, Russia, \\
    $^2$Moscow Institute of Physics and Technology (State University), Dolgoprudny, Russia,\\
    $^3$Budker Institute of Nuclear Physics, Novosibirsk, Russia,
    }\vspace{1cm}
\end{center}

\begin{abstract}

It has recently been shown that two-loop kite-type diagrams can be computed analytically in terms of iterated integrals with algebraic kernels. This result was obtained using a new integral representation for two-loop sunset subgraphs. In this paper, we have developed a similar representation for a three-loop banana integral in $d = 2-2\varepsilon$ dimensions. This answer can be generalized up to any given order in the $\varepsilon$-expansion and can be calculated numerically both below and above the threshold. We also demonstrate how this result can be used to compute more complex three-loop integrals containing the three-loop banana as a subgraph.       
\end{abstract}

\newpage

\tableofcontents{}\vspace{0.5cm}

\renewcommand{\theequation}{\thesection.\arabic{equation}}

\section{Introduction}

Computing Feynman diagrams, in
particular those with masses is one of the most important problems in modern quantum field theory. There are various methods for calculating these integrals\footnote{For a detailed overview of basic methods for computing loop Feynman integrals see \cite{smirnov2006feynman}.}, the most effective of which is the differential equations (DE) method\cite{KOTIKOV1991158,kotikov1991differential,kotikov1991differential2,remiddi1997differential,gehrmann2000differential, argeri2007feynman,henn2015lectures}. The latter is essentially based on the existence of the so-called integration by parts (IBP) relations \cite{tkachov1981theorem,chetyrkin1981integration,laporta2000high}, due to which any integral from a given family can be represented as a linear combination of a finite number of master integrals. The number of master integrals is fixed and determined by critical points of the integrand in Feynman or Baykov representation\cite{LeeCP}.    

Feynman integrals are usually expressed in terms of special functions. The multiple polylogarithms (MPLs)\cite{goncharov2,goncharov3} are proved to be the most successful here. For MPLs there are many functional dependencies and, which is important, they can be calculated numerically with high precision, see \cite{Vollinga:2004sn, Naterop:2019xaf} and references therein. The system of differential equations for the system of master integral can be solved in terms of MPLs if it can be reduced to the so-called $\ep$-form\cite{henn2013multiloop, Lee:2014ioa}, which exists only in certain cases. However, it is known that not every system of differential equations can be reduced to the $\ep$-form. In these cases, MPLs are no longer sufficient. To solve such equations, it is necessary to involve more complex functions, the simplest functions beyond multiple polylogarithms are the so-called elliptic polylogarithms (eMPLs)\cite{Beilinson:1994,Wildeshaus,Levin:1997,Levin:2007,Enriquez:2010,Brown:2011,Bloch:2013tra,Adams:2014vja,Bloch:2014qca,Adams:2015gva,Adams:2015ydq,Adams:2016xah,Remiddi:2017har,Broedel:2017kkb,Broedel:2017siw,Broedel:2018iwv,Broedel:2018qkq,Broedel:2019hyg,Broedel:2019tlz,Bogner:2019lfa,Broedel:2019kmn,Walden:2020odh,Weinzierl:2020fyx}, but wider extensions are also possible\cite{Adams:2018bsn,Adams:2018kez,Bloch:2014qca,Primo:2017ipr,Bourjaily:2017bsb,Bourjaily:2018ycu,Bourjaily:2018yfy,Bezuglov:2020ywm}.  

The purpose of this work is to generalize the functions and methods used to calculate the two-loop sunset graph from the work \cite{Bezuglov:2020ywm} to the case of the three-loop banana graph with equal masses\footnote{For other methods of calculating banana integrals, see \cite{Broedel:2019kmn,Primo:2017ipr,Bloch:2014qca,Klemm} and references therein, a similar elliptic integral also occurs when calculating the $\rho$ parameter at three loops, see \cite{Ablinger:2017bjx, Blumlein:2018aeq, Abreu:2019fgk} and references therein.}. We restrict ourselves to the case in $d = 2-2\varepsilon$ dimensions using the analogy with work \cite{Bezuglov:2020ywm} where in this case the results were more compact. We also demonstrate with a simple example that the obtained results can be used to compute three-loop diagrams which contain three-loop banana as a subgraph. The remainder of the paper is organized as follows. In section 2, we derive a new representation for three non-trivial master integrals from the three-loop banana family. Next, in section 3, we use this representation to compute a simple three-loop integral containing the three-loop banana as a subgraph. In the last section, we will draw our conclusions. Finally, in the Appendix, we will explain our notations for iterated integrals. 

\section{Banana graph}
\begin{figure}[h]
	\center{\includegraphics[width=0.4\textwidth]{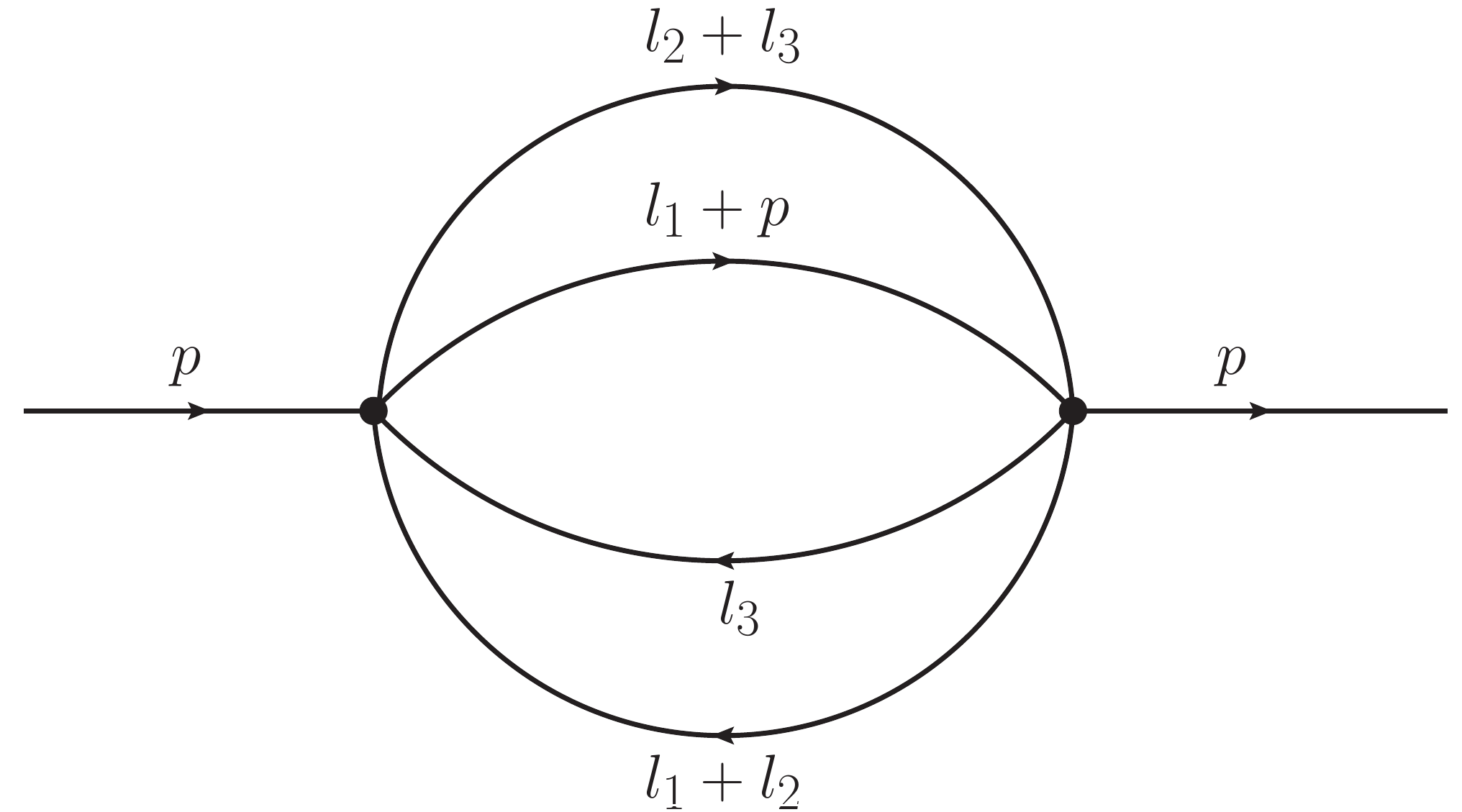}}
	\caption{Banana graph.}
	\label{bananaGraph}
\end{figure}
\begin{figure}[h]
	\center{\includegraphics[width=0.9\textwidth]{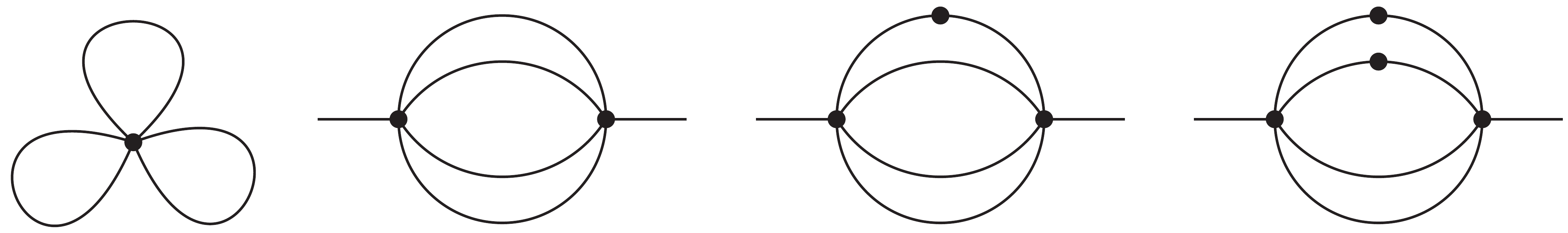}}
	\caption{Set of IBP master integrals for the three-loop banana family.}
	\label{bananaGraphLaporta}
\end{figure}

First let us define the following notation for the master integrals in the elliptic banana family, see Fig. \ref{bananaGraph}. 
\begin{equation}
j^{ban}(a_1,...,a_4)=e^{3\ep\gamma_E}\int\frac{d^dl_1d^dl_2d^dl_3}{(i\pi^{d/2})^3}\frac{1}{\left(1-l_3^2\right)^{a_1}\left(1-(l_2+l_3)^2\right)^{a_2}\left(1-(l_1+l_2)^2\right)^{a_3}\left(1-(l_1+p)^2\right)^{a_4}}
\end{equation}
with $d=2-2\ep$  and $p^2=s$.
The vector of four IBP master integrals obtained as a result of IBP reduction\cite{tkachov1981theorem,chetyrkin1981integration,laporta2000high} can be chosen in the following form:
\begin{equation}
I_{\rm IBP} = \{ j^{ban}(1,1,1,0),~j^{ban}(1,1,1,1),~ j^{ban}(2,1,1,1),~j^{ban}(2,1,2,1)\}^{\top},
\label{banLaporta}
\end{equation}
a graphical representation of these master integrals can be found in Fig. \ref{bananaGraphLaporta}. 

The first master integral is a simple constant and can be expressed analytically in the following form:
\begin{equation}
j^{ban}(1,1,1,0)=\Gamma(\ep)^3
\end{equation}

In order to find the other three master integrals we use Feynman parameter trick for pairing two pairs of propagators \footnote{For an example of using such a trick for non-elliptic cases see  \cite{effectivemass1,effectivemass2,effectivemass3,effectivemass4,effectivemass5} and references therein.} \cite{KKOVelliptic1,KKOVelliptic2,LinearReducibledEllipticFeynmanIntegrals, Bezuglov:2020ywm}, and we introduce a new family of integrals defined as($\bar{t}_{1,2}=1-t_{1,2}$) 
\begin{equation}
j^{sub}(b_1,b_2; t_1, t_2)=e^{3\ep\gamma_E}\int\frac{d^dl_1d^dl_2d^dl_3}{(i\pi^{d/2})^3}\frac{1}{\left(1-t_1l_3^2-\bar{t}_1(l_2+l_3)^2\right)^{b_1}\left(1-t_2(l_1+l_2)^2-\bar{t}_2(l_1+p)^2\right)^{b_2}}
\label{subFamily}
\end{equation}

where $t_1$ and $t_2$ are Feynman parameters which run through the unit segment($t_1,t_2 \in [0,1]$), for the convenience of the reader, we will suppress the dependence on parameters $t_1$ and $t_2$ in the following and we write instead $ j^{sub}(b_1,b_2; t_1, t_2)=j^{sub}(b_1,b_2)$. Then the three nontrivial master integrals from \eqref{banLaporta} can be expressed by integrating the integrals of the family \eqref{subFamily} over these parameters, we have 

\begin{equation}
j^{ban}(n,1,m,1)=n m \int\limits_0^1t_1^{n-1}dt_1\int\limits_0^1t_2^{m-1}dt_2 j^{sub}(n+1,m+1)
\label{intRep}
\end{equation}

Now we will use the DE method in order to evaluate the necessary integrals from the family \eqref{subFamily}. The vector of three IBP master integrals can be chosen in such a way that they can be immediately substituted into expression \eqref{intRep} without any IBP reduction

\begin{equation}
I_{\rm IBP} = \{j^{sub}(2,2),~ j^{sub}(3,2),~j^{sub}(3,3)\}^{\top}
\label{subLaporta}
\end{equation}

To evaluate these master integrals we consider their system of differential equations with respect to the variable $x$ which is associated with the old variable $s$ in the following way
\begin{equation}
s=-\frac{(xy_1-y_2)(xy_2-y_1)}{x}
\end{equation}
where we have introduced new notations
\begin{equation}
y_1=\frac{1}{\sqrt{t_1(1-t_1)}}, \qquad y_2=\frac{1}{\sqrt{t_2(1-t_2)}}.
\end{equation}

With the help of IBP identities the system of differential equations with respect to the variable $x$ can be reduced to the $\ep$-form\footnote{The subsequent reduction of system of differential equations to $\ep$-form was performed with the use of Libra \cite{LeeLibra} package and the IBP reduction was performed with the help of LiteRed \cite{LiteRed1,LiteRed2} pacadge} \cite{henn2013multiloop, Lee:2014ioa} and we have:

\begin{equation}
\label{bEq}
\frac{d \tilde{I}_{\rm canonical}}{dx}=\ep\mathcal{M}\tilde{I}_{\rm canonical}
\end{equation}
where($\alpha=\frac{y_2}{y_1}$):
\begin{equation}
\mathcal{M}=\frac{1}{x}\mathcal{M}_0+\frac{1}{x-1}\mathcal{M}_1+\frac{1}{x+1}\mathcal{M}_{-1}+\frac{1}{x-\alpha}\mathcal{M}_{\alpha}+\frac{1}{x-1/\alpha}\mathcal{M}_{1/\alpha},
\end{equation}

The particular expressions for coefficient matrices $\mathcal{M}_i$ together with transformation matrix to canonical basis $T$($I_{\rm IBP} = T \tilde{I}_{\rm canonical}$) can be found in the accompanying Mathematica notebook.
The boundary conditions for \eqref{bEq} at $x=\alpha$($s=0$) can be found by direct integration, in the Feynman parametrization we have  
\begin{equation}
j^{sub}(b_1,b_2)\Big|_{s=0}=y_1^{2-2\ep}\alpha^{2(b_2-1+\ep)}\frac{\Gamma(b_1+b_2+3\ep-3)}{\Gamma(b_1)\Gamma(b_2)}\int\limits_0^1\frac{t^{b_1-2+\ep}(1-t)^{b_2-2+\ep}dt}{\left(t+(1-t)\alpha^2\right)^{b_1+b_2-3+3\ep}}
\end{equation}
for our values of $b_1$ and $b_2$, this integral is convergent and can be easily calculated in the form of a Laurent series, the results are as follows:

\begin{equation}
j^{sub}(2,2)\Big|_{s=0}=\frac{y_1^2\alpha^2G(0;\alpha)}{\alpha^2-1}+ \mathcal{O}(\ep), 
\end{equation}
\begin{equation}
j^{sub}(3,2)\Big|_{s=0}=\frac{y_1^2\alpha^2\left(\alpha^2-1-2G(0;\alpha)\right)}{2\left(\alpha^2-1\right)^2}+ \mathcal{O}(\ep),
\end{equation}
\begin{equation}
j^{sub}(3,3)\Big|_{s=0}=\frac{y_1^2\alpha^2\left(\alpha^4-1-4\alpha^2G(0;\alpha)\right)}{4\left(\alpha^2-1\right)^3}+ \mathcal{O}(\ep).
\end{equation}
With the boundary conditions available the solution for all master integrals \eqref{subLaporta} can be found recursively in the regularization parameter $\ep$, after substituting these results into the formula \eqref{intRep} and changing variables from $t_1,~t_2$ to $y_1,~y_2$ the results for nontrivial banana master integrals will be as follows:
\begin{equation}
j^{ban}(1,1,1,1)=\int\limits_2^{\infty}\int\limits_2^{\infty}\frac{4dy_1}{y_1^2\sqrt{y_1^2-4}}\frac{4dy_2}{y_2^2\sqrt{y_2^2-4}}\frac{2 \alpha  x y_1^2 G(0;x)}{x^2-1}+ \mathcal{O}(\ep),
\label{intB1111}
\end{equation}

\begin{equation}
j^{ban}(2,1,1,1)=\int\limits_2^{\infty}\int\limits_2^{\infty}\frac{4dy_1}{y_1^2\sqrt{y_1^2-4}}\frac{4dy_2}{y_2^2\sqrt{y_2^2-4}}\left[\frac{x^2 y_1^2 \left(x^2-2 \alpha  x+1\right) G(0;x)}{\left(x^2-1\right)^3}+\frac{x y_1^2 \left(\alpha +\alpha  x^2-2 x\right)}{2 \left(x^2-1\right)^2}\right]+ \mathcal{O}(\ep),
\label{intB2111}
\end{equation}
and
\begin{equation}
j^{ban}(2,1,2,1)=\int\limits_2^{\infty}\int\limits_2^{\infty}\frac{4dy_1}{y_1^2\sqrt{y_1^2-4}}\frac{4dy_2}{y_2^2\sqrt{y_2^2-4}}\left[B^1_{33}+B^2_{33}\right]+ \mathcal{O}(\ep),
\label{intB2121}
\end{equation}
where 
\begin{equation}
B^1_{33}=\frac{x^3 y_1^2 \left(2 \alpha +2 \alpha  x^4-3 \left(\alpha ^2+1\right) x^3+8 \alpha  x^2-3 \left(\alpha ^2+1\right) x\right) G(0;x)}{\left(x^2-1\right)^5}
\end{equation}
\begin{equation}
B^2_{33}=\frac{x^2 y_1^2 \left(\alpha
   ^2+\left(\alpha ^2+1\right) x^4-12 \alpha  x^3+10 \left(\alpha ^2+1\right) x^2-12 \alpha  x+1\right)}{4 \left(x^2-1\right)^4}
\end{equation}

or in notations from section \ref{notation}:
\begin{equation}
j^{ban}(1,1,1,1)=J\left(\Psi _{-1}^1,\omega _0^x,s\right)+J\left(\Psi _1^1,\omega _0^x,s\right)+ \mathcal{O}(\ep),
\end{equation}
\begin{multline}
j^{ban}(2,1,1,1)=-\frac{1}{4} J\left(\Psi _{-3},\omega _0^x,s\right)+\frac{3}{8} J\left(\Psi _{-2},\omega _0^x,s\right)-\frac{1}{8} J\left(\Psi _{-1},\omega _0^x,s\right)+\frac{1}{8} J\left(\Psi _1,\omega
   _0^x,s\right)+\frac{3}{8} J\left(\Psi _2,\omega _0^x,s\right)+
   \\
   +\frac{1}{4} J\left(\Psi _3,\omega _0^x,s\right)-\frac{1}{4} J\left(\Psi _{-3}^1,\omega _0^x,s\right)+\frac{3}{8} J\left(\Psi _{-2}^1,\omega
   _0^x,s\right)-\frac{3}{8} J\left(\Psi _2^1,\omega _0^x,s\right)-\frac{1}{4} J\left(\Psi _3^1,\omega _0^x,s\right)-\frac{1}{4} J\left(\Psi _{-2},s\right)+
   \\
   +\frac{1}{4} J\left(\Psi
   _{-1},s\right)-\frac{1}{4} J\left(\Psi _1,s\right)-\frac{1}{4} J\left(\Psi _2,s\right)-\frac{1}{4} J\left(\Psi _{-2}^1,s\right)+\frac{1}{4} J\left(\Psi _{-1}^1,s\right)+\frac{1}{4} J\left(\Psi
   _1^1,s\right)+\frac{1}{4} J\left(\Psi _2^1,s\right)+ \mathcal{O}(\ep),
\end{multline}
\begin{multline}
j^{ban}(2,1,2,1)=\frac{3}{16} J\left(\Psi _{-5},\omega _0^x,s\right)-\frac{15}{32} J\left(\Psi _{-4},\omega _0^x,s\right)+\frac{21}{64} J\left(\Psi _{-3},\omega _0^x,s\right)-\frac{3}{128} J\left(\Psi _{-2},\omega
   _0^x,s\right)-
   \\-\frac{3}{128} J\left(\Psi _{-1},\omega _0^x,s\right)+\frac{3}{128} J\left(\Psi _1,\omega _0^x,s\right)-\frac{3}{128} J\left(\Psi _2,\omega _0^x,s\right)-\frac{21}{64} J\left(\Psi _3,\omega
   _0^x,s\right)-\frac{15}{32} J\left(\Psi _4,\omega _0^x,s\right)-
   \\
   -\frac{3}{16} J\left(\Psi _5,\omega _0^x,s\right)+\frac{3}{8} J\left(\Psi _{-5}^1,\omega _0^x,s\right)-\frac{15}{16} J\left(\Psi
   _{-4}^1,\omega _0^x,s\right)+\frac{23}{32} J\left(\Psi _{-3}^1,\omega _0^x,s\right)-\frac{9}{64} J\left(\Psi _{-2}^1,\omega _0^x,s\right)+
   \\
   +\frac{9}{64} J\left(\Psi _2^1,\omega _0^x,s\right)+\frac{23}{32}
   J\left(\Psi _3^1,\omega _0^x,s\right)+\frac{15}{16} J\left(\Psi _4^1,\omega _0^x,s\right)+\frac{3}{8} J\left(\Psi _5^1,\omega _0^x,s\right)+\frac{3}{16} J\left(\Psi _{-5}^2,\omega
   _0^x,s\right)-
   \\
   -\frac{15}{32} J\left(\Psi _{-4}^2,\omega _0^x,s\right)+\frac{21}{64} J\left(\Psi _{-3}^2,\omega _0^x,s\right)-\frac{3}{128} J\left(\Psi _{-2}^2,\omega _0^x,s\right)-\frac{3}{128}
   J\left(\Psi _{-1}^2,\omega _0^x,s\right)+\frac{3}{128} J\left(\Psi _1^2,\omega _0^x,s\right)-
   \\
   -\frac{3}{128} J\left(\Psi _2^2,\omega _0^x,s\right)-\frac{21}{64} J\left(\Psi _3^2,\omega
   _0^x,s\right)-\frac{15}{32} J\left(\Psi _4^2,\omega _0^x,s\right)-\frac{3}{16} J\left(\Psi _5^2,\omega _0^x,s\right)+\frac{3}{16} J\left(\Psi _{-4},s\right)-\frac{3}{8} J\left(\Psi
   _{-3},s\right)+
   \\
   +\frac{5}{32} J\left(\Psi _{-2},s\right)+\frac{1}{32} J\left(\Psi _{-1},s\right)-\frac{1}{32} J\left(\Psi _1,s\right)+\frac{5}{32} J\left(\Psi _2,s\right)+\frac{3}{8} J\left(\Psi
   _3,s\right)+\frac{3}{16} J\left(\Psi _4,s\right)+\frac{3}{8} J\left(\Psi _{-4}^1,s\right)-
   \\
   -\frac{3}{4} J\left(\Psi _{-3}^1,s\right)+\frac{3}{8} J\left(\Psi _{-2}^1,s\right)-\frac{3}{8} J\left(\Psi
   _2^1,s\right)-\frac{3}{4} J\left(\Psi _3^1,s\right)-\frac{3}{8} J\left(\Psi _4^1,s\right)+\frac{3}{16} J\left(\Psi _{-4}^2,s\right)-\frac{3}{8} J\left(\Psi _{-3}^2,s\right)+
   \\
   +\frac{5}{32} J\left(\Psi
   _{-2}^2,s\right)+\frac{1}{32} J\left(\Psi _{-1}^2,s\right)-\frac{1}{32} J\left(\Psi _1^2,s\right)+\frac{5}{32} J\left(\Psi _2^2,s\right)+\frac{3}{8} J\left(\Psi _3^2,s\right)+\frac{3}{16} J\left(\Psi
   _4^2,s\right)+ \mathcal{O}(\ep).
\end{multline}

\begin{figure}[h!]
	\center{\includegraphics[width=0.4\textwidth]{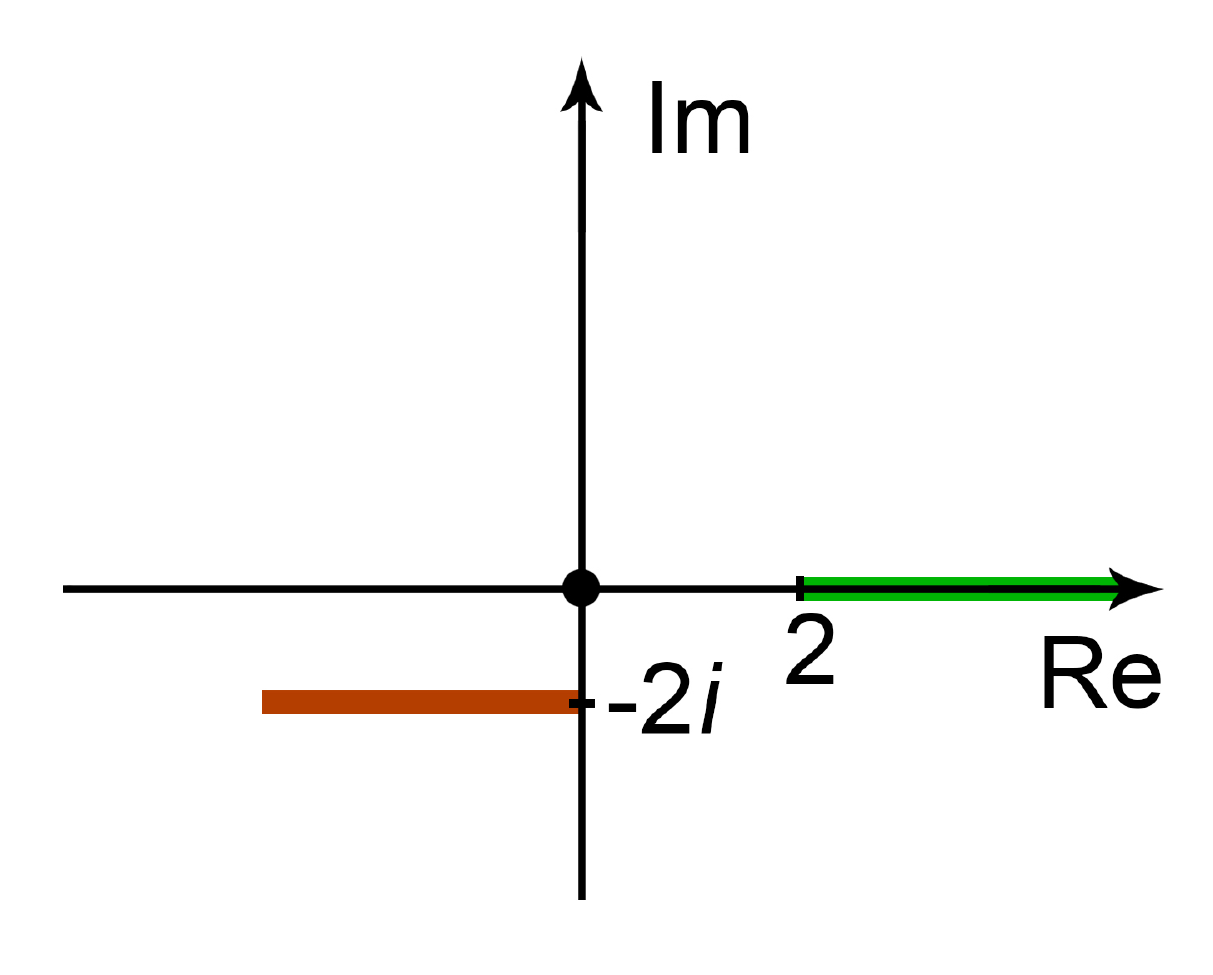}}
	\caption{Change of the integration contour, green - old integration contour, red - new integration contour. Note, that we do not just replace $y_{1,2} \rightarrow i y_{1,2}$, we also additionally deform the integration contour itself so that the integration goes to $-\infty$ instead of $-i \infty$, such deformation is possible since the integration contour lying entirely at infinity makes a zero contribution to the integral.}
	\label{intCounturs}
\end{figure}

\begin{figure}[h]
\centering
\begin{minipage}{0.45\textwidth}
\includegraphics[width=\textwidth]{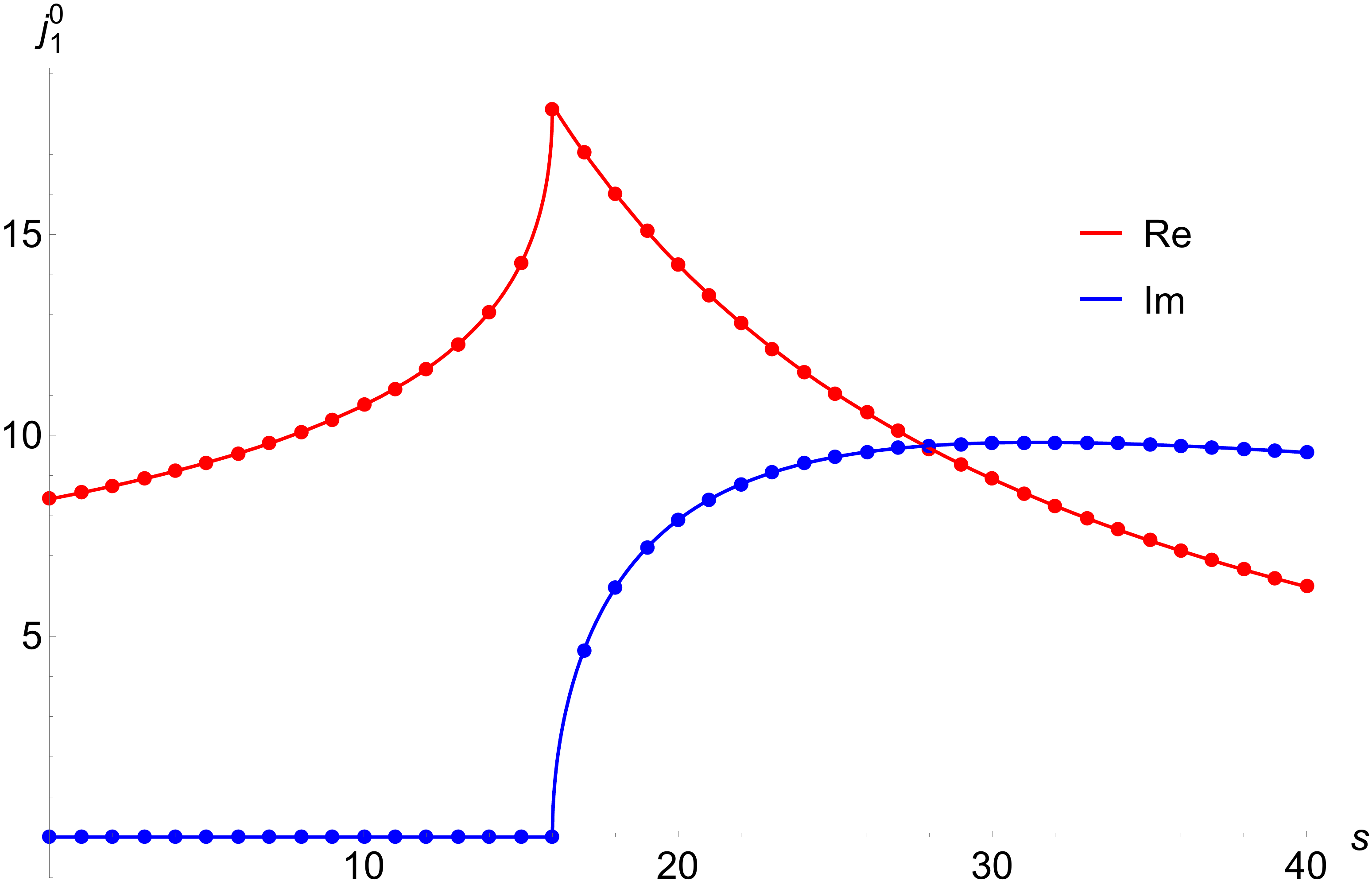}
\caption{Plot of the $\ep^0$ correction to the $j^{ban}(1,1,1,1)$ integral. The solid points represent values computed numerically with the FIESTA package\cite{Fiesta4}.}
\label{graph1}
\end{minipage}\hfill
\begin{minipage}{0.45\textwidth}
\includegraphics[width=\textwidth]{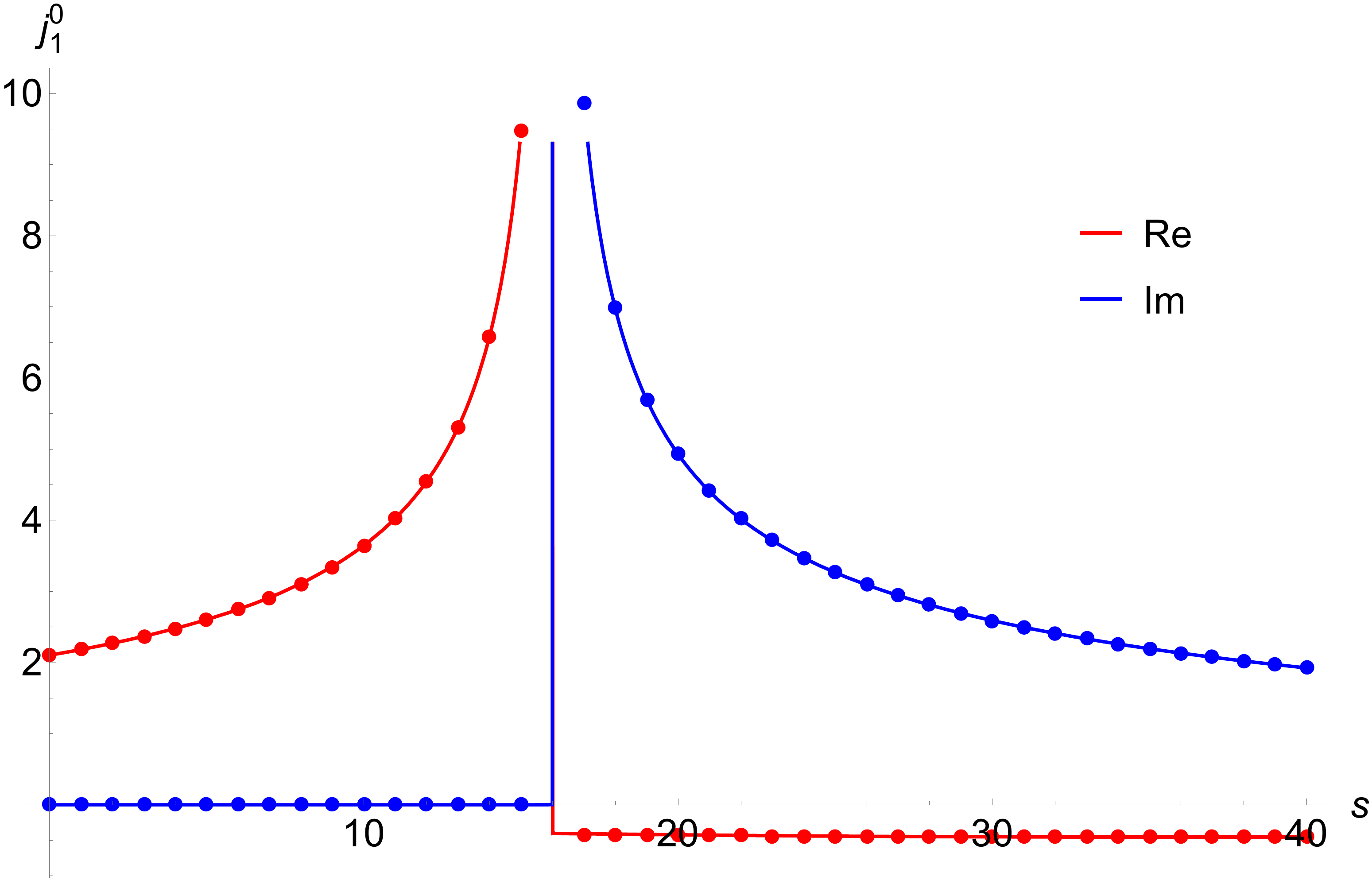}
\caption{Plot of the $\ep^0$ correction to the $j^{ban}(2,1,1,1)$ integral. The solid points represent values computed numerically with the FIESTA package\cite{Fiesta4}.}
\label{graph2}
\end{minipage}\vfill  
\begin{minipage}{0.45\textwidth}
\includegraphics[width=\textwidth]{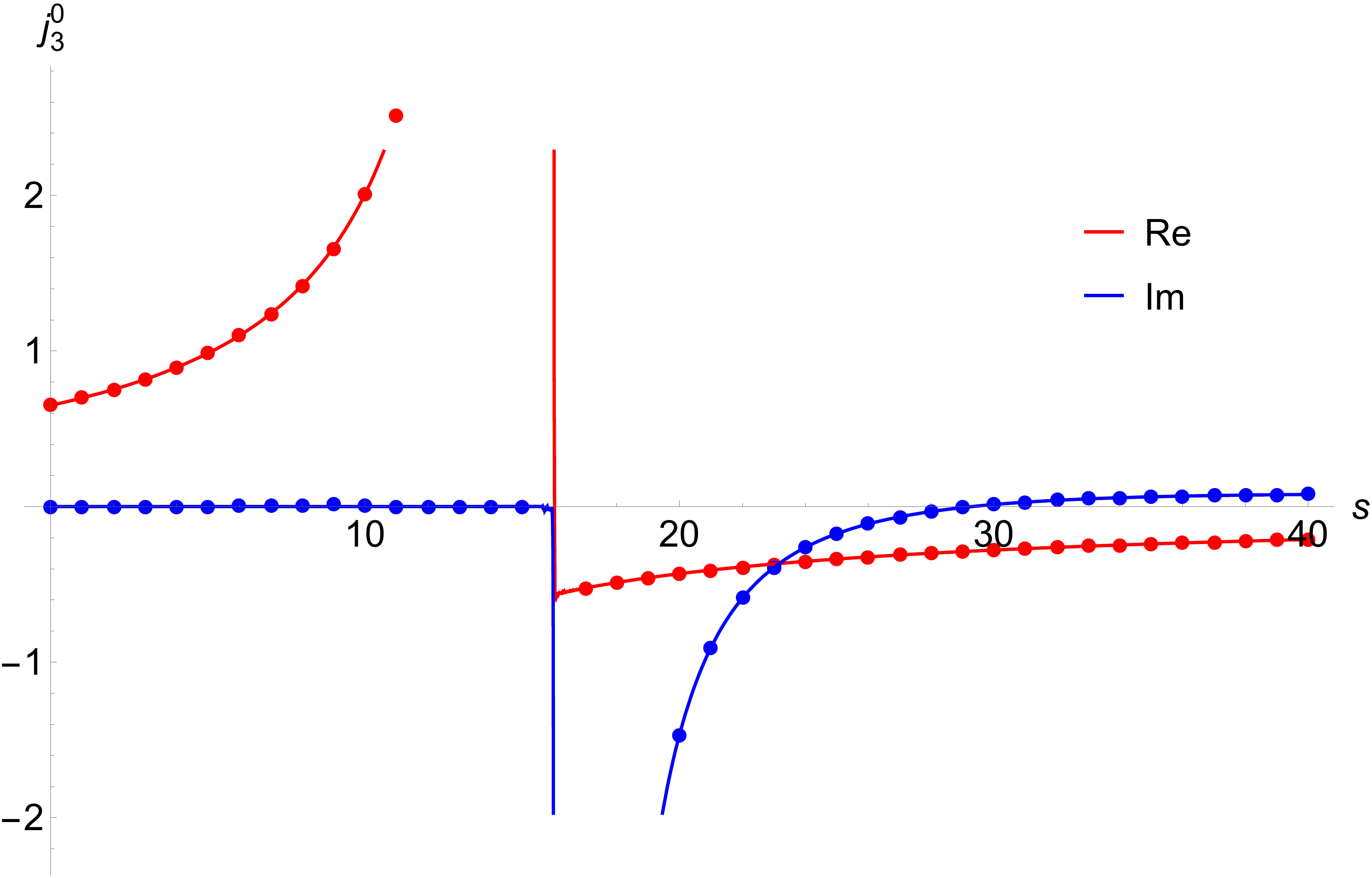}
\caption{Plot of the $\ep^0$ correction to the $j^{ban}(2,1,2,1)$ integral. The solid points represent values computed numerically with the FIESTA package\cite{Fiesta4}.}
\label{graph3}
\end{minipage}\hfill
\begin{minipage}{0.45\textwidth}
\includegraphics[width=\textwidth]{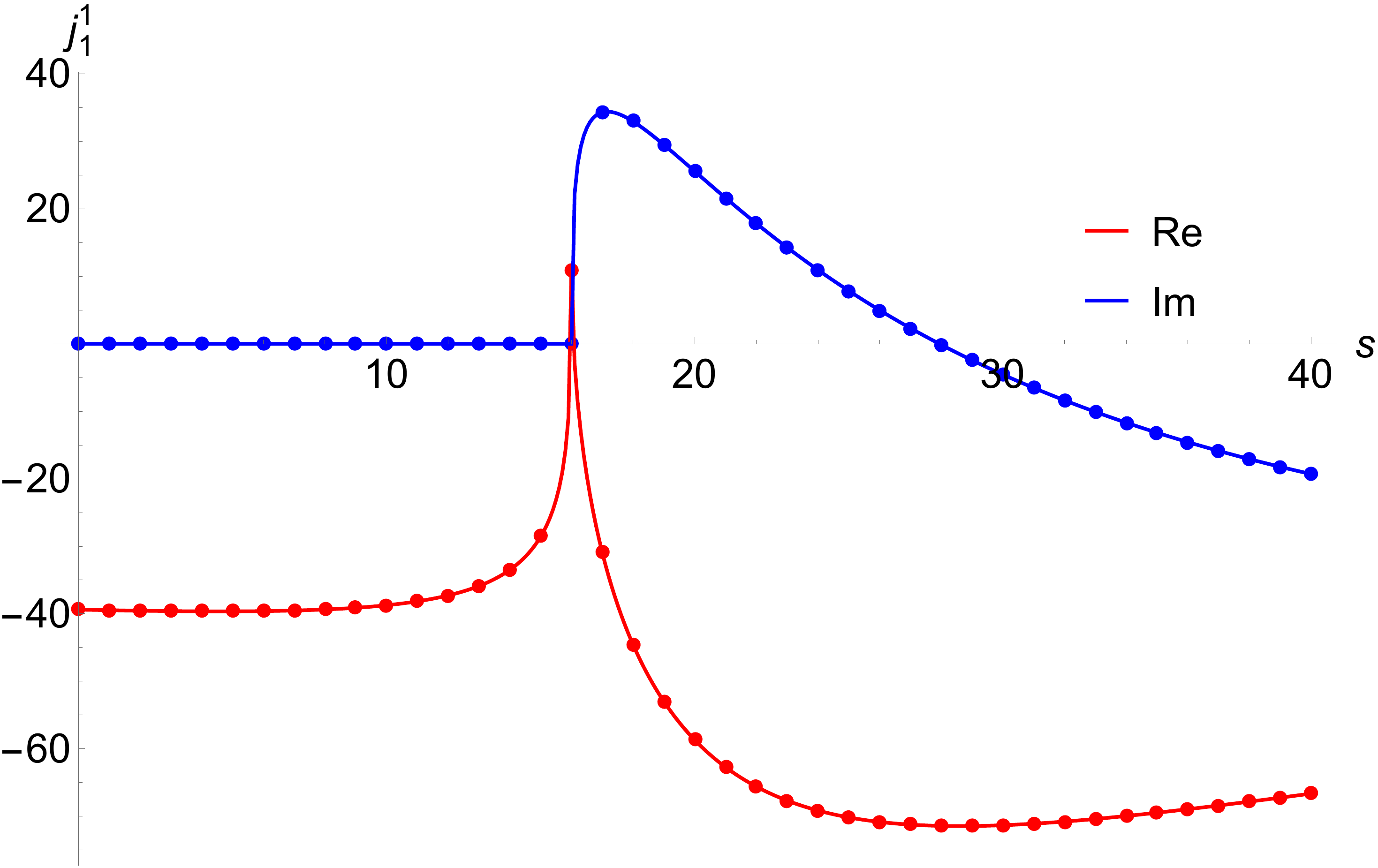}
\caption{Plot of the $\ep^1$ correction to the $j^{ban}(1,1,1,1)$ integral. The solid points represent values computed numerically with the FIESTA package\cite{Fiesta4}.}
\label{graph4}
\end{minipage}

\end{figure}

Results for master integrals \eqref{banLaporta} up to  $\mathcal{O}(\ep^2)$ corrections can be found in accompanying Mathematica file.

Integrals in equations \eqref{intB1111},\eqref{intB2111} and \eqref{intB2121} as well as higher $\ep$ corrections can be taken numerically, for this it is convenient to change the integration variables $y_{1,2} \rightarrow i y_{1,2}$ and change the contour of integration to $y_{1,2} \in [2i, -\infty]$, see Fig. \ref{intCounturs}. The example of numerical integration and their comparison with the sector decomposition method\cite{SD1,SD2,SD3,SD4,SD5,SD6,SD7} for the $\ep^0$ corrections and $\ep^1$ correction for $j^{ban}(1,1,1,1)$ integral  can be found in Fig. \ref{graph1}, \ref{graph2}, \ref{graph3}, and \ref{graph4}. 

\section{Triangle with two massive loops}
\begin{figure}[h]
	\center{\includegraphics[width=0.5\textwidth]{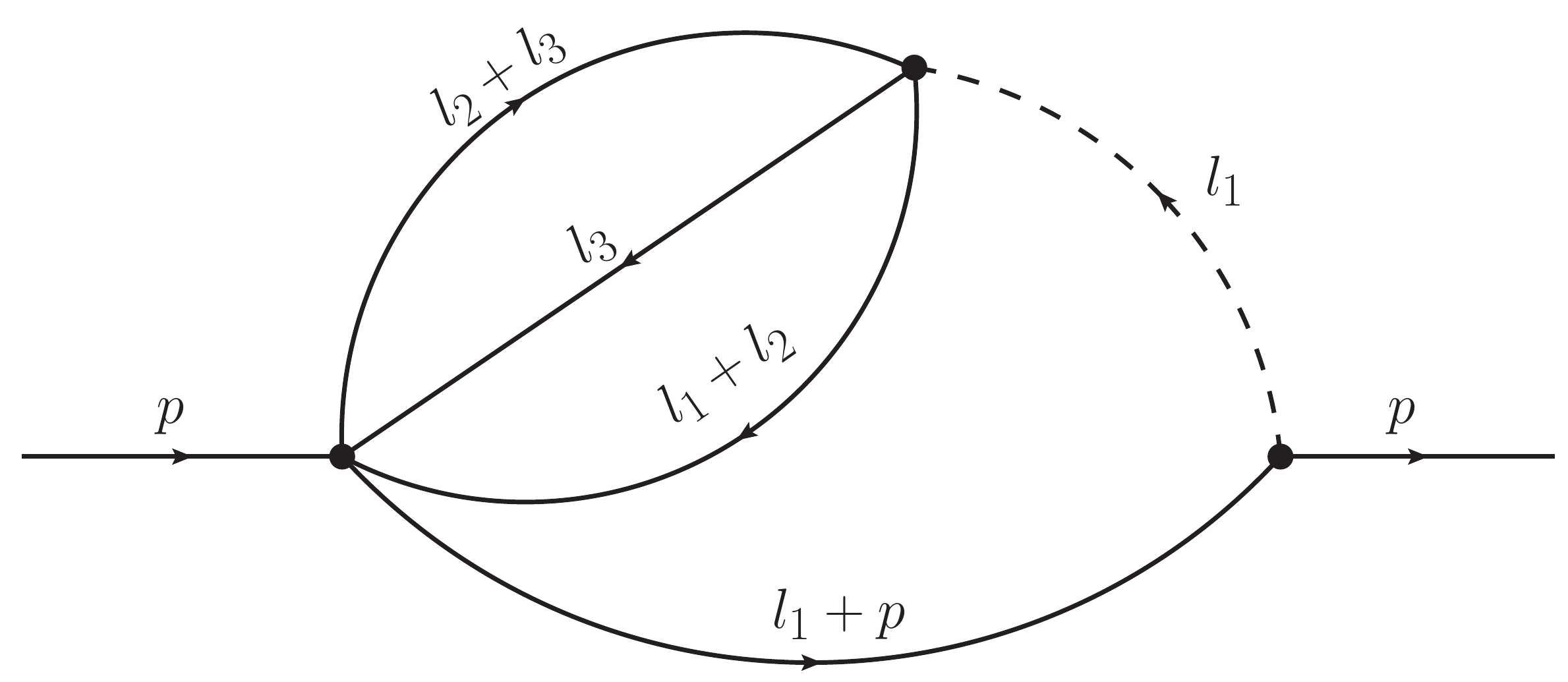}}
	\caption{Triangle with two massive loops. Dashed lines denote massless
propagators and thick lines represent massive propagators.}
	\label{Tri}
\end{figure}

\begin{figure}[h]
	\center{\includegraphics[width=0.85\textwidth]{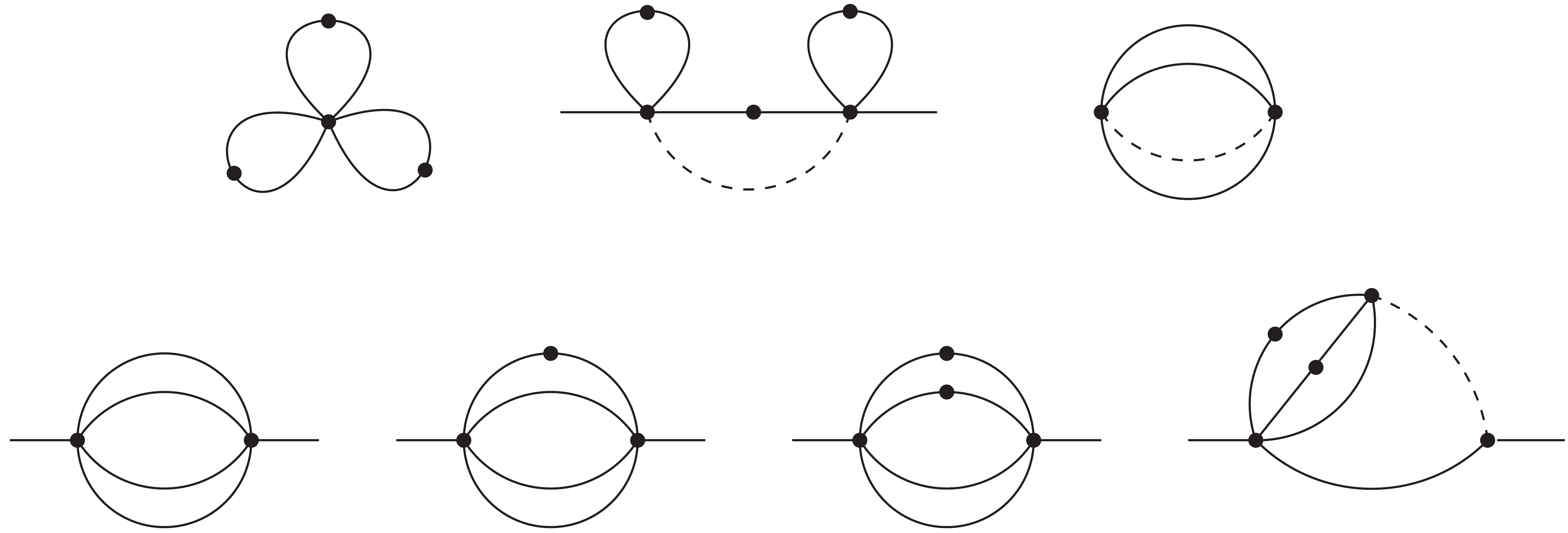}}
	\caption{Set of IBP master integrals for the family \eqref{tri1Family}. Dashed lines denote massless
propagators and thick lines represent massive propagators.}
	\label{TriLaporta}
\end{figure}

In the previous section, we obtained  an integral representation for the three-loop banana family; in this section, we will show how this representation can be used to compute more complex Feynman integrals. For this purpose, we will use the family associated with the triangle with two massive loops defined as (see Figure \ref{Tri}): 

\begin{equation}
j^{tri}(a_1,...,a_5)=\frac{e^{3\ep\gamma_E}}{(i\pi^{d/2})^3}\int\frac{d^dl_1d^dl_2d^dl_3}{\left(1-l_3^2\right)^{a_1}\left(1-(l_2+l_3)^2\right)^{a_2}\left(1-(l_1+l_2)^2\right)^{a_3}\left(1-(l_1+p)^2\right)^{a_4}\left(-l_1^2\right)^{a_5}}
\label{tri1Family}
\end{equation}
with $d=4-2\ep$  and $p^2=s$.
The vector of seven IBP master integrals obtained as a result of IBP reduction\cite{tkachov1981theorem,chetyrkin1981integration,laporta2000high} together
with dimension recurrence relations\cite{dimrecurrence} can be chosen in the following form:
\begin{align}
I_{\rm IBP} = \{j^{tri}(0,2,2,2,0),~j^{tri}(0,2,2,2,1),~j^{tri}(1,1,1,0,1),~j^{ban}(1,1,1,1),\nonumber \\~ j^{ban}(2,1,1,1),~j^{ban}(2,1,2,1),~j^{tri}(2,2,1,1,1)\}^{\top}
\label{triLaporta}
\end{align}
a graphical representation of these master integrals can be found in Fig. \ref{TriLaporta}. 

The first and third master integrals are simple constants and can be written as 

\begin{equation}
j^{tri}(0,2,2,2,0)=\Gamma(\ep)^3
\end{equation}
and
\begin{equation}
j^{tri}(1,1,1,0,1) = \frac{1}{\ep^3}+\frac{15}{4\ep^2}+\frac{65+2\pi^2}{8\ep}+\left(\frac{135}{16}-\zeta_3+\frac{45}{8}\zeta_2+\frac{81}{4}S_2\right)+ \mathcal{O}(\ep)
\end{equation}
where $S_2= \frac{4}{9\sqrt{3}}\text{Cl}_2\left(\frac{\pi}{3}\right)$. And the integrals $j^{ban}(1,1,1,1), ~ j^{ban}(2,1,1,1)$, and $j^{ban}(2,1,2,1)$ were found in the previous section. 

To evaluate the remaining master integrals we consider their system of differential equations with respect
to $p^2=s$. Using balance transformations of \cite{Lee:2014ioa} via the package \cite{LeeLibra} the latter can be reduced to the following $A+B\ep$
form:

\begin{equation}
\frac{d \tilde{I}_{\rm canonical}}{ds}=\mathcal{A}\tilde{I}_{\rm canonical}
\end{equation}
with
\begin{equation}
\mathcal{A}=\frac{1}{s}\mathcal{A}_0+\frac{1}{s-1}\mathcal{A}_1+\frac{1}{s-4}\mathcal{A}_{4}+\frac{1}{s-16}\mathcal{A}_{16},
\end{equation}
and
\begin{equation}
\mathcal{A}_0=
\left(
\begin{array}{ccccccc}
 0 & 0 & 0 & 0 & 0 & 0 & 0 \\
 0 & \epsilon  & 0 & 0 & 0 & 0 & 0 \\
 0 & 0 & 0 & 0 & 0 & 0 & 0 \\
 0 & 0 & 0 & -3 \epsilon -1 & 4 (4 \epsilon +1) & 0 & 0 \\
 0 & 0 & 0 & \frac{1}{4} (-3 \epsilon -1) & 4 \epsilon +1 & 0 & 0 \\
 \frac{\epsilon }{8} & 0 & 0 & -\frac{3}{32} (2 \epsilon +1) & \frac{5}{8} (2 \epsilon +1) & -\epsilon -1 & 0 \\
 0 & 0 & 0 & \frac{5}{12} (4 \epsilon +1) & -\frac{5}{3} (4 \epsilon +1) & 0 & \epsilon  \\
\end{array}
\right),
\end{equation}

\begin{equation}
\mathcal{A}_1=
\left(
\begin{array}{ccccccc}
 0 & 0 & 0 & 0 & 0 & 0 & 0 \\
 -\epsilon  & -2 \epsilon  & 0 & 0 & 0 & 0 & 0 \\
 0 & 0 & 0 & 0 & 0 & 0 & 0 \\
 0 & 0 & 0 & 0 & 0 & 0 & 0 \\
 0 & 0 & 0 & 0 & 0 & 0 & 0 \\
 0 & 0 & 0 & 0 & 0 & 0 & 0 \\
 0 & 0 & 0 & \frac{1}{6} (-11 \epsilon -2) & \frac{5}{3} (4 \epsilon +1) & 0 & -2 \epsilon  \\
\end{array}
\right),
\end{equation}

\begin{equation}
\mathcal{A}_4=
\left(
\begin{array}{ccccccc}
 0 & 0 & 0 & 0 & 0 & 0 & 0 \\
 0 & 0 & 0 & 0 & 0 & 0 & 0 \\
 0 & 0 & 0 & 0 & 0 & 0 & 0 \\
 0 & 0 & 0 & 0 & 0 & 0 & 0 \\
 0 & 0 & 0 & \frac{1}{4} (3 \epsilon +1) & -2 (3 \epsilon +1) & 3 (3 \epsilon +1) & 0 \\
 0 & 0 & 0 & \frac{1}{8} (2 \epsilon +1) & -2 \epsilon -1 & \frac{3}{2} (2 \epsilon +1) & 0 \\
 0 & 0 & 0 & 0 & 0 & 0 & 0 \\
\end{array}
\right),
\end{equation}
\begin{equation}
\mathcal{A}_{16}=
\left(
\begin{array}{ccccccc}
 0 & 0 & 0 & 0 & 0 & 0 & 0 \\
 0 & 0 & 0 & 0 & 0 & 0 & 0 \\
 0 & 0 & 0 & 0 & 0 & 0 & 0 \\
 0 & 0 & 0 & 0 & 0 & 0 & 0 \\
 0 & 0 & 0 & 0 & 0 & 0 & 0 \\
 -\frac{\epsilon }{8} & 0 & 0 & \frac{1}{32} (-2 \epsilon -1) & \frac{3}{8} (2 \epsilon +1) & -\frac{3}{2} (2 \epsilon +1) & 0 \\
 0 & 0 & 0 & 0 & 0 & 0 & 0 \\
\end{array}
\right).
\end{equation}
And the elements of the canonical basis  $\tilde{I}_{\rm canonical}=\{I_1, ..., I_7\}^{\top}$ are related to the elements of the IBP basis \eqref{TriLaporta} as
\begin{align}
I_1 & = \ep^2 j^{tri}(0,2,2,2,0),\\
I_2 & = s\ep^2 j^{tri}(0,2,2,2,1),\\
I_3 & = j^{tri}(1,1,1,0,1),\\
I_4 & =(1+3\ep)(1+4\ep) j^{ban}(1,1,1,1),\\
I_5 & =(1+3\ep)j^{ban}(2,1,1,1),\\
I_6 & =j^{ban}(2,1,2,1),\\
I_7 & =\frac{(1+3\ep)(1+4\ep)}{s-1} \left(\frac{5}{12} j^{ban}(1,1,1,1) - s(1-2\ep)j^{tri}(2,2,1,1,1)\right).\\
\end{align}

Having obtained the differential system in this
form it is easy to see, that the solution for required master integrals $j^{tri}(0,2,2,2,1)$ and $j^{tri}(2,2,1,1,1)$ can be obtained recursively in the regularization parameter $\ep$ similarly to what one typically does for differential systems reduced to $\ep$-form. Of course, of greatest interest is the solution for the integral $j^{tri}(2,2,1,1,1)$ which can be written through the J-functions from the Appendix 
\begin{multline}
j^{tri}(2,2,1,1,1) = \frac{s-1}{12s}\Bigg[
-5 J\left(\Lambda _0,\zeta _{0,-1}^s,s\right)-5 J\left(\Lambda _0,\zeta _{0,1}^s,s\right)+5 J\left(\Lambda _0,\zeta _{1,-1}^s,s\right)+5 J\left(\Lambda _0,\zeta
   _{1,1}^s,s\right)-
   \\
   -5 J\left(\Lambda _1,\zeta _{0,-1}^s,s\right)+5 J\left(\Lambda _1,\zeta _{0,1}^s,s\right)+5 J\left(\Lambda _1,\zeta _{1,-1}^s,s\right)-5
   J\left(\Lambda _1,\zeta _{1,1}^s,s\right)+5 J\left(\Lambda _0,\omega _{0,-1}^s,s\right)
   -
   \\
   -5 J\left(\Lambda _0,\omega _{0,1}^s,s\right)-5 J\left(\Lambda _0,\omega
   _{1,-1}^s,s\right)+5 J\left(\Lambda _0,\omega _{1,1}^s,s\right)+5 J\left(\Lambda _1,\omega _{0,-1}^s,s\right)+5 J\left(\Lambda _1,\omega _{0,1}^s,s\right)-
   \\
   -5
   J\left(\Lambda _1,\omega _{1,-1}^s,s\right)-5 J\left(\Lambda _1,\omega _{1,1}^s,s\right)+\frac{15}{2} J\left(\Lambda _0,\zeta _{0,-1}^s,\omega
   _0^x,s\right)+\frac{15}{2} J\left(\Lambda _0,\zeta _{0,1}^s,\omega _0^x,s\right)-
   \\
   -\frac{15}{2} J\left(\Lambda _0,\zeta _{1,-1}^s,\omega
   _0^x,s\right)-\frac{15}{2} J\left(\Lambda _0,\zeta _{1,1}^s,\omega _0^x,s\right)+\frac{15}{2} J\left(\Lambda _1,\zeta _{0,-1}^s,\omega
   _0^x,s\right)-\frac{15}{2} J\left(\Lambda _1,\zeta _{0,1}^s,\omega _0^x,s\right)-
   \\
   -\frac{15}{2} J\left(\Lambda _1,\zeta _{1,-1}^s,\omega
   _0^x,s\right)+\frac{15}{2} J\left(\Lambda _1,\zeta _{1,1}^s,\omega _0^x,s\right)-5 J\left(\Lambda _0,\eta _{0,-1}^s,\omega _0^x,s\right)+5 J\left(\Lambda
   _0,\eta _{0,1}^s,\omega _0^x,s\right)+
   \\
   +5 J\left(\Lambda _0,\eta _{1,-1}^s,\omega _0^x,s\right)-5 J\left(\Lambda _0,\eta _{1,1}^s,\omega _0^x,s\right)-5
   J\left(\Lambda _1,\eta _{0,-1}^s,\omega _0^x,s\right)-5 J\left(\Lambda _1,\eta _{0,1}^s,\omega _0^x,s\right)+
   \\
   +5 J\left(\Lambda _1,\eta _{1,-1}^s,\omega
   _0^x,s\right)+5 J\left(\Lambda _1,\eta _{1,1}^s,\omega _0^x,s\right)-\frac{5}{2} J\left(\Lambda _0,\omega _{0,-1}^s,\omega _0^x,s\right)+\frac{5}{2}
   J\left(\Lambda _0,\omega _{0,1}^s,\omega _0^x,s\right)+
\\   
   +\frac{5}{2} J\left(\Lambda _0,\omega _{1,-1}^s,\omega _0^x,s\right)-\frac{5}{2} J\left(\Lambda _0,\omega
   _{1,1}^s,\omega _0^x,s\right)-5 J\left(\Lambda _1,\omega _{0,-1}^s,\omega _0^x,s\right)-5 J\left(\Lambda _1,\omega _{0,1}^s,\omega _0^x,s\right)+
   \\
   +4
   J\left(\Lambda _1,\omega _{1,-1}^s,\omega _0^x,s\right)+4 J\left(\Lambda _1,\omega _{1,1}^s,\omega _0^x,s\right)+\frac{5 J\left(\Psi _{-1}^1,\omega
   _0^x,s\right)}{s-1}+\frac{5 J\left(\Psi _1^1,\omega _0^x,s\right)}{s-1}\Bigg] + \mathcal{O}(\ep)
\end{multline}
For reference, we also present the result for the second master integral, which can be expressed in terms of usual MPLs:
\begin{multline}
j^{tri}(0,2,2,2,1)=-\frac{G(1,s)}{s \ep^2}+\frac{2 G(1,1,s)-G(0,1,s)}{s \ep}-\frac{\pi ^2 G(1,s)}{4 s}-\frac{G(0,0,1,s)}{s}+
\\
+\frac{2
   G(0,1,1,s)}{s}+\frac{2 G(1,0,1,s)}{s}-\frac{4 G(1,1,1,s)}{s}+ \mathcal{O}(\ep)
\end{multline}

Note, that with the use of the presented procedure we can have as many terms in $\ep$ expansion of considered master integrals as required.

\section{Conclusion}
\label{Conclusion}
In this paper, we have obtained a new representation for the three-loop equal-mass banana graph in $d=2-2\ep$ dimensions. These results are written in terms of new functions defined as iterated integrals with algebraic kernels. These functions have already been used earlier in \cite{Bezuglov:2020ywm} to compute the two-loop sunset diagram as well as the massive kite diagram. Our work can be seen as a straightforward generalization of techniques from \cite{Bezuglov:2020ywm} to the three-loop case. The obtained representation for the three-loop banana graph can be used to calculate some more complex three-loop graphs, we have illustrated the last statement by using the example of the triangle with two massive loops. The analytical results for the three-loop banana can be numerically calculated with good accuracy both above and below the threshold and are in agreement with the sector decomposition method \cite{SD1,SD2,SD3,SD4,SD5,SD6,SD7} as implemented in \cite{Fiesta4}. The result for a triangle with two massive loops was verified numerically only below the threshold and its analytical continuation above it will be
the subject of our future research.  

\section*{Acknowledgements}
I would like to thank A.I. Onishchenko for interesting
and stimulating discussions as well as for general guidance in writing this work.
The work was supported by the Foundation for the Advancement of Theoretical Physics and Mathematics "BASIS".

\appendix

\section{Notation for iterated integrals}\label{notation}
The obtained in the paper results for banana and triangle with two massive loops diagrams can be conveniently expressed in terms of iterated integrals with algebraic kernels of the form:

\begin{equation}
J(\Psi,\omega_1^s,\ldots ,\omega_n^s,\omega_1^x,\ldots ,\omega_n^x,\omega_1^{\alpha},\ldots ,\omega_m^{\alpha},\omega_1^{y_1},\ldots ,\omega_l^{y_1};s)
\end{equation}
where $\Psi$ is some 2-form in $y_1$ and $y_2$ integrated in the limits $y_{1,2} \in [2, \infty]$ and $\omega^s$,$\omega^x$,$\omega^{\alpha}$, and $\omega^{y_1}$, are some 1-form in $s$, $x$, $\alpha$, and $y_1$ respectively and J-function form iterated integrals in these 1-forms. For example, we have
\begin{equation}
J\left(
\Psi, \frac{dx}{x-1}, \frac{d\alpha}{\alpha-1}, \frac{dy_1}{y_1}; s
\right) = \int_2^{\infty}\int_2^{\infty}\frac{dy_1}{y_1^2\sqrt{y_1^2-4}}\frac{dy_2}{y_2^2\sqrt{y_2^2-4}}\int_0^x\frac{dx'}{x'-1}\int_0^{\alpha}\frac{d\alpha'}{\alpha'-1}\int_0^{y_1}\frac{dy_1'}{y_1'}\, .
\end{equation}

In general, iterated integrals in our results contain the following $\Psi$ and $\Lambda$  2-forms ($J_y=\frac{4}{y^2\sqrt{y^2-4}}$): 

\begin{equation}
\Psi_{\pm n}=\frac{y_1^2J_{y_1}J_{y_2}dy_1dy_2}{(x\mp 1)^n}, \qquad \Psi_{\pm n}^m=\frac{y_1^2\alpha^mJ_{y_1}J_{y_2}dy_1dy_2}{(x\mp 1)^n}, \qquad \Lambda_m = y_1^2\alpha^mJ_{y_1}J_{y_2}dy_1dy_2
\end{equation}

and  $\omega$, $\zeta$ and $\eta$ 1-forms:
\begin{equation}
\omega_a^x = \frac{dx}{x-a}, \qquad \omega_b^{\alpha} = \frac{d\alpha}{\alpha-b}, \qquad \omega_c^{y_1} = \frac{dy_1}{y_1-c}, \qquad \omega_a^s = \frac{ds}{s-a},
\end{equation}

\begin{equation}
\omega_{a,b}^s = \frac{ds}{(s-a)(x-b)}, \qquad \zeta_{a,b}^s = \frac{ds}{(s-a)(x-b)^2}, \qquad \eta_{a,b}^s = \frac{ds}{(s-a)(x-b)^3}. \qquad 
\end{equation}

Where we tried to choose notations so that they, if possible, coincide with notations from \cite{Bezuglov:2020ywm}.

\bibliographystyle{ieeetr}
\bibliography{litr}

\begin{thebibliography}{10}

\bibitem{smirnov2006feynman}
V.~Smirnov, {\em Feynman Integral Calculus}.
\newblock Springer Berlin Heidelberg, 2006.

\bibitem{KOTIKOV1991158}
A.~Kotikov, ``Differential equations method. new technique for massive feynman
  diagram calculation,'' {\em Physics Letters B}, vol.~254, no.~1, pp.~158 --
  164, 1991.

\bibitem{kotikov1991differential}
A.~Kotikov, ``Differential equation method. the calculation of n-point feynman
  diagrams,'' {\em Physics Letters B}, vol.~267, no.~1, pp.~123--127, 1991.

\bibitem{kotikov1991differential2}
A.~Kotikov, ``Differential equations method: the calculation of vertex-type
  feynman diagrams,'' {\em Physics Letters B}, vol.~259, no.~3, pp.~314--322,
  1991.

\bibitem{remiddi1997differential}
E.~Remiddi, ``Differential equations for feynman graph amplitudes,'' {\em Il
  Nuovo Cimento A (1971-1996)}, vol.~110, no.~12, pp.~1435--1452, 1997.

\bibitem{gehrmann2000differential}
T.~Gehrmann and E.~Remiddi, ``Differential equations for two-loop four-point
  functions,'' {\em Nuclear Physics B}, vol.~580, no.~1-2, pp.~485--518, 2000.

\bibitem{argeri2007feynman}
M.~Argeri and P.~Mastrolia, ``Feynman diagrams and differential equations,''
  {\em International Journal of Modern Physics A}, vol.~22, no.~24,
  pp.~4375--4436, 2007.

\bibitem{henn2015lectures}
J.~M. Henn, ``Lectures on differential equations for feynman integrals,'' {\em
  Journal of Physics A: Mathematical and Theoretical}, vol.~48, no.~15,
  p.~153001, 2015.

\bibitem{tkachov1981theorem}
F.~V. Tkachov, ``A theorem on analytical calculability of 4-loop
  renormalization group functions,'' {\em Physics Letters B}, vol.~100, no.~1,
  pp.~65--68, 1981.

\bibitem{chetyrkin1981integration}
K.~G. Chetyrkin and F.~V. Tkachov, ``Integration by parts: the algorithm to
  calculate $\beta$-functions in 4 loops,'' {\em Nuclear Physics B}, vol.~192,
  no.~1, pp.~159--204, 1981.

\bibitem{laporta2000high}
S.~Laporta, ``High-precision calculation of multiloop feynman integrals by
  difference equations,'' {\em International Journal of Modern Physics A},
  vol.~15, no.~32, pp.~5087--5159, 2000.

\bibitem{LeeCP}
R.~N. Lee and A.~A. Pomeransky, ``{Critical points and number of master
  integrals},'' {\em JHEP}, vol.~11, p.~165, 2013.

\bibitem{goncharov2}
A.~B. Goncharov, ``Multiple polylogarithms, cyclotomy and modular complexes,''
  {\em Mathematical Research Letters}, vol.~5, pp.~497--516, 1998.

\bibitem{goncharov3}
A.~B. Goncharov, ``Multiple polylogarithms and mixed tate motives,'' {\em arXiv
  preprint math/0103059}, 2001.

\bibitem{Vollinga:2004sn}
J.~Vollinga and S.~Weinzierl, ``{Numerical evaluation of multiple
  polylogarithms},'' {\em Comput. Phys. Commun.}, vol.~167, p.~177, 2005.

\bibitem{Naterop:2019xaf}
L.~Naterop, A.~Signer, and Y.~Ulrich, ``{handyG \textemdash{}Rapid numerical
  evaluation of generalised polylogarithms in Fortran},'' {\em Comput. Phys.
  Commun.}, vol.~253, p.~107165, 2020.

\bibitem{henn2013multiloop}
J.~M. Henn, ``Multiloop integrals in dimensional regularization made simple,''
  {\em Physical review letters}, vol.~110, no.~25, p.~251601, 2013.

\bibitem{Lee:2014ioa}
R.~N. Lee, ``{Reducing differential equations for multiloop master
  integrals},'' {\em JHEP}, vol.~04, p.~108, 2015.

\bibitem{Beilinson:1994}
A.~Beilinson and A.~Levin, ``Elliptic polylogarithms,'' {\em Proc. of Symp. in
  Pure Mathematics}, vol.~55, pp.~126--196, 1994.

\bibitem{Wildeshaus}
J.~Wildeshaus {\em Lect. Notes Math.}, vol.~1650, 1997.

\bibitem{Levin:1997}
A.~Levin, ``Elliptic polylogarithms: An analytic theory,'' {\em Compositio
  Mathematica}, vol.~106, no.~3, p.~267–282, 1997.

\bibitem{Levin:2007}
A.~Levin and G.~Racinet, ``{Towards multiple elliptic polylogarithms},'' 2007.

\bibitem{Enriquez:2010}
B.~Enriquez, ``Elliptic associators,'' 2012.

\bibitem{Brown:2011}
F.~C.~S. Brown and A.~Levin, ``Multiple elliptic polylogarithms,'' 2013.

\bibitem{Bloch:2013tra}
S.~Bloch and P.~Vanhove, ``{The elliptic dilogarithm for the sunset graph},''
  {\em J. Number Theor.}, vol.~148, pp.~328--364, 2015.

\bibitem{Adams:2014vja}
L.~Adams, C.~Bogner, and S.~Weinzierl, ``{The two-loop sunrise graph in two
  space-time dimensions with arbitrary masses in terms of elliptic
  dilogarithms},'' {\em J. Math. Phys.}, vol.~55, no.~10, p.~102301, 2014.

\bibitem{Bloch:2014qca}
S.~Bloch, M.~Kerr, and P.~Vanhove, ``{A Feynman integral via higher normal
  functions},'' {\em Compos. Math.}, vol.~151, no.~12, pp.~2329--2375, 2015.

\bibitem{Adams:2015gva}
L.~Adams, C.~Bogner, and S.~Weinzierl, ``{The two-loop sunrise integral around
  four space-time dimensions and generalisations of the Clausen and Glaisher
  functions towards the elliptic case},'' {\em J. Math. Phys.}, vol.~56, no.~7,
  p.~072303, 2015.

\bibitem{Adams:2015ydq}
L.~Adams, C.~Bogner, and S.~Weinzierl, ``{The iterated structure of the
  all-order result for the two-loop sunrise integral},'' {\em J. Math. Phys.},
  vol.~57, no.~3, p.~032304, 2016.

\bibitem{Adams:2016xah}
L.~Adams, C.~Bogner, A.~Schweitzer, and S.~Weinzierl, ``{The kite integral to
  all orders in terms of elliptic polylogarithms},'' {\em J. Math. Phys.},
  vol.~57, no.~12, p.~122302, 2016.

\bibitem{Remiddi:2017har}
E.~Remiddi and L.~Tancredi, ``{An Elliptic Generalization of Multiple
  Polylogarithms},'' {\em Nucl. Phys.}, vol.~B925, pp.~212--251, 2017.

\bibitem{Broedel:2017kkb}
J.~Broedel, C.~Duhr, F.~Dulat, and L.~Tancredi, ``{Elliptic polylogarithms and
  iterated integrals on elliptic curves. Part I: general formalism},'' {\em
  JHEP}, vol.~05, p.~093, 2018.

\bibitem{Broedel:2017siw}
J.~Broedel, C.~Duhr, F.~Dulat, and L.~Tancredi, ``{Elliptic polylogarithms and
  iterated integrals on elliptic curves II: an application to the sunrise
  integral},'' {\em Phys. Rev.}, vol.~D97, no.~11, p.~116009, 2018.

\bibitem{Broedel:2018iwv}
J.~Broedel, C.~Duhr, F.~Dulat, B.~Penante, and L.~Tancredi, ``{Elliptic symbol
  calculus: from elliptic polylogarithms to iterated integrals of Eisenstein
  series},'' {\em JHEP}, vol.~08, p.~014, 2018.

\bibitem{Broedel:2018qkq}
J.~Broedel, C.~Duhr, F.~Dulat, B.~Penante, and L.~Tancredi, ``{Elliptic Feynman
  integrals and pure functions},'' {\em JHEP}, vol.~01, p.~023, 2019.

\bibitem{Broedel:2019hyg}
J.~Broedel, C.~Duhr, F.~Dulat, B.~Penante, and L.~Tancredi, ``{Elliptic
  polylogarithms and Feynman parameter integrals},'' {\em JHEP}, vol.~05,
  p.~120, 2019.

\bibitem{Broedel:2019tlz}
J.~Broedel and A.~Kaderli, ``{Functional relations for elliptic
  polylogarithms},'' {\em J. Phys.}, vol.~A53, no.~24, p.~245201, 2020.

\bibitem{Bogner:2019lfa}
C.~Bogner, S.~Müller-Stach, and S.~Weinzierl, ``{The unequal mass sunrise
  integral expressed through iterated integrals on $\overline{\mathcal
  M}_{1,3}$},'' {\em Nucl. Phys.}, vol.~B954, p.~114991, 2020.

\bibitem{Broedel:2019kmn}
J.~Broedel, C.~Duhr, F.~Dulat, R.~Marzucca, B.~Penante, and L.~Tancredi, ``{An
  analytic solution for the equal-mass banana graph},'' {\em JHEP}, vol.~09,
  p.~112, 2019.

\bibitem{Walden:2020odh}
M.~Walden and S.~Weinzierl, ``{Numerical evaluation of iterated integrals
  related to elliptic Feynman integrals},'' 2020.

\bibitem{Weinzierl:2020fyx}
S.~Weinzierl, ``{Modular transformations of elliptic Feynman integrals},''
  2020.

\bibitem{Adams:2018bsn}
L.~Adams, E.~Chaubey, and S.~Weinzierl, ``{Planar Double Box Integral for Top
  Pair Production with a Closed Top Loop to all orders in the Dimensional
  Regularization Parameter},'' {\em Phys. Rev. Lett.}, vol.~121, no.~14,
  p.~142001, 2018.

\bibitem{Adams:2018kez}
L.~Adams, E.~Chaubey, and S.~Weinzierl, ``{Analytic results for the planar
  double box integral relevant to top-pair production with a closed top
  loop},'' {\em JHEP}, vol.~10, p.~206, 2018.

\bibitem{Primo:2017ipr}
A.~Primo and L.~Tancredi, ``{Maximal cuts and differential equations for
  Feynman integrals. An application to the three-loop massive banana graph},''
  {\em Nucl. Phys.}, vol.~B921, pp.~316--356, 2017.

\bibitem{Bourjaily:2017bsb}
J.~L. Bourjaily, A.~J. McLeod, M.~Spradlin, M.~von Hippel, and M.~Wilhelm,
  ``{Elliptic Double-Box Integrals: Massless Scattering Amplitudes beyond
  Polylogarithms},'' {\em Phys. Rev. Lett.}, vol.~120, no.~12, p.~121603, 2018.

\bibitem{Bourjaily:2018ycu}
J.~L. Bourjaily, Y.-H. He, A.~J. Mcleod, M.~Von~Hippel, and M.~Wilhelm,
  ``{Traintracks through Calabi-Yau Manifolds: Scattering Amplitudes beyond
  Elliptic Polylogarithms},'' {\em Phys. Rev. Lett.}, vol.~121, no.~7,
  p.~071603, 2018.

\bibitem{Bourjaily:2018yfy}
J.~L. Bourjaily, A.~J. McLeod, M.~von Hippel, and M.~Wilhelm, ``{Bounded
  Collection of Feynman Integral Calabi-Yau Geometries},'' {\em Phys. Rev.
  Lett.}, vol.~122, no.~3, p.~031601, 2019.

\bibitem{Bezuglov:2020ywm}
M.~A. Bezuglov, A.~I. Onishchenko, and O.~L. Veretin, ``{Massive kite diagrams
  with elliptics},'' {\em Nucl. Phys. B}, vol.~963, p.~115302, 2021.

\bibitem{Klemm}
A.~Klemm, C.~Nega, and R.~Safari, ``{The $l$-loop Banana Amplitude from GKZ
  Systems and relative Calabi-Yau Periods},'' {\em JHEP}, vol.~04, p.~088,
  2020.

\bibitem{Ablinger:2017bjx}
J.~Ablinger, J.~Bl\"umlein, A.~De~Freitas, M.~van Hoeij, E.~Imamoglu, C.~G.
  Raab, C.~S. Radu, and C.~Schneider, ``{Iterated Elliptic and Hypergeometric
  Integrals for Feynman Diagrams},'' {\em J. Math. Phys.}, vol.~59, no.~6,
  p.~062305, 2018.

\bibitem{Blumlein:2018aeq}
J.~Bl\"umlein, A.~De~Freitas, M.~Van~Hoeij, E.~Imamoglu, P.~Marquard, and
  C.~Schneider, ``{The $\rho$ parameter at three loops and elliptic
  integrals},'' {\em PoS}, vol.~LL2018, p.~017, 2018.

\bibitem{Abreu:2019fgk}
S.~Abreu, M.~Becchetti, C.~Duhr, and R.~Marzucca, ``{Three-loop contributions
  to the $\rho$ parameter and iterated integrals of modular forms},'' {\em
  JHEP}, vol.~02, p.~050, 2020.

\bibitem{effectivemass1}
J.~Fleischer, A.~V. Kotikov, and O.~L. Veretin, ``{The Differential equation
  method: Calculation of vertex type diagrams with one nonzero mass},'' {\em
  Phys. Lett.}, vol.~B417, pp.~163--172, 1998.

\bibitem{effectivemass2}
J.~Fleischer, A.~V. Kotikov, and O.~L. Veretin, ``{Analytic two loop results
  for selfenergy type and vertex type diagrams with one nonzero mass},'' {\em
  Nucl. Phys.}, vol.~B547, pp.~343--374, 1999.

\bibitem{effectivemass3}
J.~Fleischer, M.~{\relax Yu}. Kalmykov, and A.~V. Kotikov, ``{Two loop
  selfenergy master integrals on-shell},'' {\em Phys. Lett.}, vol.~B462,
  pp.~169--177, 1999.
\newblock [Erratum: Phys. Lett.B467,310(1999)].

\bibitem{effectivemass4}
B.~A. Kniehl and A.~V. Kotikov, ``{Calculating four-loop tadpoles with one
  non-zero mass},'' {\em Phys. Lett.}, vol.~B638, pp.~531--537, 2006.

\bibitem{effectivemass5}
B.~A. Kniehl and A.~V. Kotikov, ``{Counting master integrals:
  integration-by-parts procedure with effective mass},'' {\em Phys. Lett.},
  vol.~B712, pp.~233--234, 2012.

\bibitem{KKOVelliptic1}
B.~A. Kniehl, A.~V. Kotikov, A.~Onishchenko, and O.~Veretin, ``{Two-loop sunset
  diagrams with three massive lines},'' {\em Nucl. Phys.}, vol.~B738,
  pp.~306--316, 2006.

\bibitem{KKOVelliptic2}
B.~A. Kniehl, A.~V. Kotikov, A.~I. Onishchenko, and O.~L. Veretin, ``{Two-loop
  diagrams in non-relativistic QCD with elliptics},'' {\em Nucl. Phys.},
  vol.~B948, p.~114780, 2019.

\bibitem{LinearReducibledEllipticFeynmanIntegrals}
M.~Hidding and F.~Moriello, ``{All orders structure and efficient computation
  of linearly reducible elliptic Feynman integrals},'' {\em JHEP}, vol.~01,
  p.~169, 2019.

\bibitem{LeeLibra}
R.~N. Lee, ``{Libra: a package for transformation of differential systems for
  multiloop integrals},'' 12 2020.

\bibitem{LiteRed1}
R.~N. Lee, ``{Presenting LiteRed: a tool for the Loop InTEgrals REDuction},''
  2012.

\bibitem{LiteRed2}
R.~N. Lee, ``{LiteRed 1.4: a powerful tool for reduction of multiloop
  integrals},'' {\em J. Phys. Conf. Ser.}, vol.~523, p.~012059, 2014.

\bibitem{Fiesta4}
A.~V. Smirnov, ``{FIESTA4: Optimized Feynman integral calculations with GPU
  support},'' {\em Comput. Phys. Commun.}, vol.~204, pp.~189--199, 2016.

\bibitem{SD1}
T.~Binoth and G.~Heinrich, ``{An automatized algorithm to compute infrared
  divergent multiloop integrals},'' {\em Nucl. Phys.}, vol.~B585, pp.~741--759,
  2000.

\bibitem{SD2}
T.~Binoth and G.~Heinrich, ``{Numerical evaluation of multiloop integrals by
  sector decomposition},'' {\em Nucl. Phys.}, vol.~B680, pp.~375--388, 2004.

\bibitem{SD3}
T.~Binoth and G.~Heinrich, ``{Numerical evaluation of phase space integrals by
  sector decomposition},'' {\em Nucl. Phys.}, vol.~B693, pp.~134--148, 2004.

\bibitem{SD4}
G.~Heinrich, ``{Sector Decomposition},'' {\em Int. J. Mod. Phys.}, vol.~A23,
  pp.~1457--1486, 2008.

\bibitem{SD5}
C.~Bogner and S.~Weinzierl, ``{Resolution of singularities for multi-loop
  integrals},'' {\em Comput. Phys. Commun.}, vol.~178, pp.~596--610, 2008.

\bibitem{SD6}
C.~Bogner and S.~Weinzierl, ``{Blowing up Feynman integrals},'' {\em Nucl.
  Phys. Proc. Suppl.}, vol.~183, pp.~256--261, 2008.

\bibitem{SD7}
T.~Kaneko and T.~Ueda, ``{A Geometric method of sector decomposition},'' {\em
  Comput. Phys. Commun.}, vol.~181, pp.~1352--1361, 2010.

\bibitem{dimrecurrence}
O.~V. Tarasov, ``{Connection between Feynman integrals having different values
  of the space-time dimension},'' {\em Phys. Rev.}, vol.~D54, pp.~6479--6490,
  1996.

\end{thebibliography}

\end{document}